\documentstyle[twoside,fleqn]{article}
\newcommand{\bls}[1]{\renewcommand{\baselinestretch}{#1}}

\def\noi{\noindent}

\def\nq{\hspace{-1em}}
\def\nqq{\hspace{-2em}}
\def\nhq{\hspace{-0.5em}}

\def\cm{\hspace{1cm}}
\def\inch{\hspace{1in}}
\def\yy{\\[5pt]}
\def\al{&\nhq}
\def\eql{\al =\al}
\def\nnn{\nonumber\\ \lal }
\def\lal{&&\nqq {}}               
\def\nn{\nonumber\\ {}}

\newcommand{\gs}{\sigma}
\newcommand{\gt}{\theta}
\def\vp{\varphi}

\def\DAL{\raisebox{-1.6pt}{\large $\Box$}\,}

\def\const{{\rm const}}
\def\eq{Eq.\,}
\def\eqs{Eqs.\,}
\newcommand{\arctg}{\arctan}

\newcommand{\ch}{\cosh}
\newcommand{\sh}{\sinh}
\newcommand{\th}{\tanh}
\newcommand{\cth}{\coth}
\newcommand{\e}{{\rm e}}
\newcommand{\Arch}{\mathop{\rm Arcosh}\nolimits}

\newcommand{\jheads}[2]{\markboth{\protect\small\it #1}{\protect\small\it #2}}
\newcommand{\Acknow}[1]{\subsection*{Acknowledgement} #1}
\newcommand{\Title}[1]{\noindent {\Large #1} \\}
\newcommand{\Author}[2]{\noindent{\large\bf #1}\\[2ex]\noindent{\it #2}\\}
\newcommand{\Abstract}[1]{\vskip 2mm \begin{center}
        \parbox{16.4cm}{\small\noi #1} \end{center}\medskip}
\newcommand{\PACS}[1]{\begin{center}{\small PACS: #1}\end{center}}
\newcommand{\email}[2]{\footnotetext[#1]{e-mail: #2}}

\def\beq{\begin{equation}}
\def\eeq{\end{equation}}
\def\bear{\begin{eqnarray}}
\def\ear{\end{eqnarray}}
\def\bearr{\begin{eqnarray} \lal}
\def\earn{\nonumber \end{eqnarray}}

\bls{1.03}
\jheads{V.M. Zhuravlev, S.V. Chervon and D.Yu. Shabalkin}
{Effective Chiral Model of a Plane-Symmetric
Gravitational Field.}

\begin{document}
\twocolumn[

\bigskip

\Title{EFFECTIVE CHIRAL MODEL OF A PLANE-SYMMETRIC \yy
GRAVITATIONAL FIELD: PROPERTIES AND EXACT SOLUTIONS}

\Author{ V.M. Zhuravlev${}^1$, S.V. Chervon and D.Yu. Shabalkin}
{Ulyanovsk State University,
42 Leo Tolstoy St., Ulyanovsk 432700, Russia}

\Abstract
{An effective chiral model of a plane-symmetric gravitational
field is considered. Isometries of the target space of the
model are described and exact solutions corresponding to the
isometric ansatz method are obtained. New exact solutions
are found using the method of functional parameters.
The solutions obtained are B\"acklund transforms of solutions of
the d'Alembert equation to those of the Einstein equations.
                  }

] 

\email 1 {zhuravl@themp.univ.simbirsk.su}

\section{Introduction}

One of the most interesting nonlinear field models is the chiral
(bosonic) nonlinear sigma model which has a lot of possibilities to be
investigated by geometrical methods \cite{ch97m}.
It was pointed out that a certain class of Einstein gravitational field
equations can be represented as the dynamical equations of a nonlinear
sigma model (NSM) with a special choice of a target space
\cite{matmis67,chemus1,chemus2}:
\beq\label{DEnsm}
{1\over\sqrt {|g|}}\partial_i(\sqrt {|g|} \varphi_{A}^{,i})-
 \Gamma_{C,AB} \varphi_{,i}^B \varphi_{,k}^C g^{ik}=0,
\eeq
which follow from the action
\beq\label{Snsm}
{\cal S}= \int\limits_{\cal M} \sqrt{|g|} d^4x \lbrace\frac{1}{2}
 h_{AB}(\vp) \vp^A_{,i} \vp^B_{,k} g^{ik} \rbrace.
\eeq
The notations corresponded to those in \cite{ch97gc}.
Using this representation, one can apply the isometric ansatz method
\cite{ivanov83-38} with the aim to find exact solutions
describing gravitational fields. On the other hand, a direct analysis of
the transformed Einstein equations is possible as well.

In the present article isometries of the target space of the
effective chiral model of a plane-symmetric gravitational field
are considered and exact solutions corresponding to the isometric ansatz
method are obtained. New exact solutions for the
model under consideration are found by the method of functional
parameters.  The solutions obtaines are the results of B\"acklund
transformations of the D'Alembert equation to the Einstein equations.

\section{Chiral effective model}

The effective chiral model of a plane-symmetric gravitational field
\bear \label{int}
     ds^2 \eql A(dx^1)^2+2B dx^1 dx^2+C(dx^2)^2 \nnn
              \cm\cm    - D[ (dx^3)^2-(dx^4)^2]
\ear
is considered. The metric coefficients $A,B,C,D$ depend only on $x^3,x^4$.

The vacuum Einstein equation
 \beq\label{VEE}
 R_{ik}=0
 \eeq
 for the space-time (\ref{int}), as was shown in
\cite{chemus1}, may be written as a four-component nonlinear sigma model
for the fields
$\varphi^1=\psi$, $\varphi^2=\theta$, $\varphi^3=\chi$, $\varphi^4=\phi$,
defined on the two-dimentional space-time
\beq\label{space}
dS^2=(dx^3)^2-(dx^4)^2.
\eeq
The metric of the target space for the effective chiral model can be
reduced to the form \cite{chemus1}
\beq\label{chispace}
h_{IK}={\e^\psi}     \pmatrix {-1&0&0&1\cr
                                 0&1&0&0\cr
                                 0&0& \sh^2\theta&0\cr
                                 1&0&0&0\cr}.
\eeq
The metric coefficients should be connected with the chiral field by
\bear\label{MetCoef}
A \eql -\e^\psi({\rm cos}\chi \sinh\theta+\cosh\theta),\nn
B \eql \e^\psi \sin\chi \sinh\theta,\nn
C \eql \e^\psi (\cos\chi \sinh\theta-\cosh\theta),\nn
D \eql \e^\phi.
\ear
The vacuum Einstein equations (\ref{VEE}) may be written as the dynamic
equations   \cite{chemus1}
\bear\label{NSMequat}
\Delta \e^\psi \eql 0, \nn
 \Delta\theta+(\psi_3\theta_3-\psi_4\theta_4)-{1\over2}(\chi^2_3-\chi^2_4)
  \sinh 2\theta\eql 0,\nn
 \nq \ \Delta\chi+2(\theta_3\chi_3-\theta_4\chi_4)\coth\theta
 +(\psi_3\chi_3-\psi_4\chi_4) \eql 0,\nn
 \Delta\phi+{1\over2}(\psi^2_3-\psi^2_4)           \cm  \inch \lal \nn
            -{1\over2}(\chi^2_3-\chi^2_4)\sinh^2\theta
 - {1\over2}(\theta^2_3-\theta^2_4)  \eql 0, \nnn \\
  \Delta\equiv\partial_{33}-\partial_{44}. \inch \lal
\ear

To finish our construction, it is necessary to check the
reversibility of the transformation (\ref{MetCoef}). The determinant of
the transformation Jacobian is equal to
\beq\label{Jac}
     \det J=\sinh\theta.
\eeq
So one should assume the only restriction following from the
condition (\ref{Jac}) to transfer the analysis from
(\ref{VEE}) to (\ref{NSMequat}).

There are a lot of verified methods of studying nonlinear sigma models.
In this work we try to apply to the effective chiral model analysis the
method of an isometric ansatz based on its symmetry properties.

\section{Isometries of the effective chiral model}

The allowed isometric motions in the chiral space and their connection
with the space-time symmetry will be examined.

The isometries of the target space (\ref{chispace}) can be obtained
by solving the Killing equations
\beq\label{KE}
\zeta_{A;B}+\zeta_{B;A}=0.
\eeq
(\ref{KE}) results in the linearly independent Killing vectors
\bearr\label{KILL1}
\zeta_1=(0,\cos \chi,-\sin \chi \cth \theta,0),\\ \lal
\zeta_2=(0,\sin \chi,\cos \chi \cth \theta,0),\\ \lal
\zeta_3=(0,0,1,0),\quad \zeta_4=(0,0,0,1).
\ear

The Killing vectors of the space-time (\ref{space}) are
\[
\xi_1^i=\delta^i_4,\qquad
\xi_2 ^i=\delta^i_3,\qquad
\xi^i_3=x^3\delta^i_4-x^4\delta^i_3.
\]

To study the symmetry properties of the model, we use the
isometric ansatz introduced by G.G.~Ivanov \cite{ivanov83-38}
\beq\label{ansatz}
        \xi_a^i\varphi_i^A=K_a^{\alpha}\zeta_{\alpha}^A.
\eeq
Here $\varphi_i^A$ mean the derivatives of the field $\varphi^A$ in $x^i$
and the constants $K_a^{\alpha}$ are the expansion coefficients.

Consider a subgroup of the basic space-time isometry
associated with $\vec\xi_1$. The ansatz may be written as
\bearr         \label{ansatz1}
\psi_4=0,\qquad \phi_4=d, \qquad
                    \theta_4=-\tilde a\sin\tilde \chi,\nnn
\tilde \chi_4=-\tilde a \cth \theta \cos \tilde \chi +m;\yy  \lal
\tilde a=\sqrt{a^2+b^2},\qquad
                        \tilde \chi = \chi + \alpha,\qquad     a=K_1^1,\nnn
b=K_1^2,\qquad  \alpha=\arctan(b/a),\qquad m=K_1^3,\nnn
d=K_1^4.
\ear
Solutions will be sought in the case $m = 0$, $d\not =0$, $\tilde a \not = 0$.

A solution to the ansatz may be written in the form
\bearr              \label{SolAn}
\psi=\psi(x^3),\cm         \phi=d\cdot x^4 + f(x^3),\nnn
\theta=\Arch\biggl\{\frac{\sqrt{\tilde a X^2}}{\tilde a X}
\ch\left\{\tilde a\left[ x^4-B(x^3)\right]\right\}\biggr\},\nnn
\chi=\arccos\left( \frac{1}{\tilde a X \sh \theta}\right)-\alpha,\nnn
             X=X(x^3).
\ear

\section{Exact solution}

The ansatz solution singles out some class of chiral fields
for which it is possible to solve the field equations (\ref{NSMequat}).
To this end it is necessary to determine the parametric
functions $X=X(x^3)$, $B=B(x^3)$ and the fields $\psi(x^3)$, $\phi(x^3,x^4)$.
Substitution of the ansatz solution to the effective chiral model equation
leads to separation of variables connected with a dependence of the
functions $X$ and $B$ on only $x^3$. A necessary condition for the existence
of solutions to the effective chiral model in the form (\ref{SolAn}) is
$X(x^3)=\alpha$ and $B(x^3)=\beta x^3$ ($\alpha,\ \beta = \const$).
Further research has shown that the sufficient condition makes
narrower the class of possible effective chiral model solution, which is
rather narrow itself.  So the only case when the ansatz solution
(\ref{SolAn}) satisfies the effective chiral model equation
(\ref{NSMequat}) is as follows:
\bearr       \label{SolAnfil}
\psi=c,\cm             \phi=d\cdot x^4 + p x^3,\nnn
\theta=\Arch\left\{  1/\sqrt{\tilde a }
                \ch\left(\tilde a\left[ x^4-x^3\right]\right)\right\},\nnn
\chi=\arccos\left[ 1/(\tilde a  \sh \theta)\right]-\alpha,
\ear
$c$ and $p$ being constants.

The metric coefficients found according to (\ref{MetCoef}) are
\bearr                                                 \label{SolMet}
A=-\frac{1}{\tilde a}\e^{-c}
\biggl[     \frac{a}{\tilde a}+\sqrt{{\tilde a}^2+1}         \nnn
\quad\ \times  \biggl\{\frac{b}{\tilde a}|\sh[\tilde a(x^4-x^3)]|
                 +\ch[\tilde a(x^4-x^3)]    \biggr\}    \biggr],\nnn
B=\frac{1}{\tilde a}\e^c
      \biggl[ \frac{a}{\tilde a} \sqrt{{\tilde a}^2+1}
               |\sh^2 [\tilde a(x^4-x^3)]|-\frac{b}{\tilde a} \biggr],\nnn
C=-\frac{1}{\tilde a}\e^{-c}
\biggl[ \frac{a}{\tilde a}
         +\sqrt{{\tilde a}^2+1}\nnn
\quad\   \times\biggl\{\frac{b}{\tilde a}|\sh [\tilde a(x^4-x^3)]|
   -\ch[\tilde a(x^4-x^3)]   \biggr\} \biggr],\nnn
D=\e^{p[(d/p)x^4+x^3]}.
\ear
The solution is a gravitational pulse of an unchanged shape,
moving at the speed of light.

This solution cannot be generated by the inverse scattering problem
method (ISPM) formulated by Belinski and Zakharov \cite{belzak78}.

\section{Method of functional parameter. The first class of exact
solutions}

Other way of studying \eqs (\ref{NSMequat}) may be presented as follows.
Solutions of (\ref{NSMequat}) will be sought in the form
\bearr\label{Sol0}
   \gt=\gt(\xi),\qquad \chi=\chi(\xi),
   \qquad \phi=\phi(\xi), \qquad   \psi=\ln \xi \nnn
\ear
where $\xi=\xi(z,t)$ is a functional parameter satisfying
\beq\label{Solxi}
  \DAL\xi=0,
\eeq
according to the first equation of (\ref{NSMequat}).

Substitution of (\ref{Sol0}) to the other equation of (\ref{NSMequat}),
leads to the set of ordinary differential equations
\bear  \label{ODU}
  \ddot\gt+\frac{1}{\xi}\dot\gt-\frac{1}{2}\dot\chi^2\sh2\gt \eql 0,\nn
  \ddot\chi+\frac{1}{\xi}\dot\chi+{2}\dot\chi\dot\gt\cth\gt \eql 0,\nn
  \ddot\phi+\frac{1}{2\xi^2}-\frac{1}{2}\dot\gt^2
               -\frac{1}{4}\dot\chi^2\sh^2\gt \eql 0,
\ear
for the functions $\gt$, $\chi$ and $\phi$ of the
parameter $\xi$ treated as an independent variable.
Here and henceforth
\[
     \dot{\phi_a}=\frac{d}{d\xi}\phi_a, \quad
     \ddot{\phi_a}=\frac{d^2}{d\xi^2}\phi_a,\ldots;~a=1,2,3,4.
\]
\eqs (\ref{ODU}) are exactly integrated and the solution
 may be written as
\bearr\label{SolF}
   \gt_{\pm}=\ln
   \biggl\{\frac{a}{2}\left[\left(\frac{\xi}{\xi_0}\right)^k+
                \left(\frac{\xi}{\xi_0}\right)^{-k}\right]\nnn  \cm
   \pm\sqrt{\frac{a^2}{4}\left[\left(\frac{\xi}{\xi_0}\right)^{2k}+
\left(\frac{\xi}{\xi_0}\right)^{-2k}\right]+\frac{a^2}{2} - 1}\biggr\},\nnn
   \chi_{\pm}=
\chi_0\pm\arctg\left\{\frac{k}{c}\frac{(\xi/\xi_0)^k-
   (\xi/\xi_0)^{-k}}{(\xi/\xi_0)^{k}+(\xi/\xi_0)^{-k}}\right\},\\ \lal
\phi_{\pm}
   =\phi_0+\phi_1\xi-(\frac{k^2}{2}-2)(\ln\xi-1)        \nnn \inch \cm
   -\frac{c}{4}
   \xi\int^\xi\frac{\chi_{\pm}(\xi')}{\xi'}d\xi', \nnn
\label{Sapq}
    \quad      a=\sqrt{\frac{k^2+c^2}{k^2}},
\ear
where $\phi_0,\ \phi_1,\ \chi_0,\ \xi_0,\  k,\ c$ are arbitrary real
constants.

The above pairs of solutions
$\{\gt_{\pm}(\xi)$, $\chi_{\pm}(\xi)$, $\phi_{\pm}(\xi)$, $\psi(\xi)\}$
of (\ref{ODU}) correspond to each solution
$\xi=\xi(z,t)=f(z-t)+g(z+t)$ of \eq (\ref{Solxi})
( $f(z-t)$ and $g(z+t)$ are arbitrary twice differentiable functions).
The connection (\ref{MetCoef}) between the metric coefficients  and
the chiral fields allows one to construct solutions of the Einstein
equations (\ref{VEE}) for the space-time (\ref{int}).

The condition $\xi>0$ is necessary for the field $\phi$ to be a real
function.

The transformations (\ref{MetCoef}), (\ref{NSMequat}), (\ref{SolF})
may be interpretated as a B\"acklund transformation of \eq (\ref{Solxi}) to
(\ref{VEE}) for the space-time (\ref{int}) since $\xi$ is a functional
parameter.

It is necessary to note that the B\"acklund transformation possesses
an obvious functional form. Therefore in this case no additional
construction is needed for obtaining the solutions, as it was, for example,
in the case of the ISPM of Belinski-Zakharov \cite{belzak78}. The same
shortcoming of the ISPM does not allow one to formulate in a simple way the
conditions of coincidence between the Belinski-Zakharov family of solutions
and those obtained here.

The possibility of formulating the initial and bo\-und\-ary-value problems
for the gravitational field and constructihg exact solutions is one more
advantage of this method over the ISPM.

\section{Method of functional parameter. Second class of exact solutions}

Other family of solutions which cannot be reduced to the above family,
may be obtained under the assumptions
\beq\label{Sol1}
\nq   \gt=\gt(\xi), \quad\ \chi=\chi(\xi), \quad\ \phi=\phi(\xi),
     \quad\ \psi=\const.
\eeq
Equations for $\gt(\xi)$, $\chi(\xi)$ and $\phi(\xi)$ are written in this
case in the following way:
\bear              \label{ODU1}
  \ddot\gt-\frac{1}{2}\dot\chi^2\sh2\gt  \eql  0,\nn
\ddot\chi+{2}\dot\chi\dot\gt\cth\gt      \eql  0, \nn
\ddot\phi-\frac{1}{2}\dot\gt^2-\frac{1}{4}\dot\chi^2\sh^2\gt \eql  0.
\ear
        Their solutions are constructed in a way similar to
        that described above. They are
\bearr       \label{SolF1}
 \nq  \gt_{\pm}
   =\ln\left\{a\ch k(\xi{-}\xi_0)
                        \pm\sqrt{a^2\ch^2 k(\xi{-}\xi_0) - 1}\right\},
                 \nnn
 \nq   \chi_{\pm}=\int\frac{cd\xi}{\sh^2\gt_{\pm}}=
       \chi_0\pm\arctg\left\{\frac{k}{c}\th [ k(\xi{-}\xi_0) ]\right\},
          \nnn
 \nq  \phi_{\pm}=\phi_0+\phi_1\xi+\frac{k^2}{2}\xi^2-\frac{c}{4}
   \e^{\zeta}\int\limits^\zeta \e^{-\zeta'}\chi_{\pm}(\zeta')d\zeta'.\nnn
\ear
The constant $a$ is given by (\ref{Sapq}).

The difference between this family of solutions and the previous one is
in the substitution of the parameter $\xi$ to $\gt$ and $\chi$
instead of $\zeta=\ln\xi$ and in the addition to $\phi$ of a term depending
on $\xi^2$.

A comparison of the solutions (\ref{SolF1}) and (\ref{SolAnfil}) shows
that the solution obtained using the isometric ansatz method
does not belong to the family (\ref{SolF1}) for any value of the
arbitrary constants and for any form of the function $\xi(x^3,x^4)$.

\section{Generalization of the method of functional parameter}

The idea of introducing a functional parameter may be generalized as follows.
Consider one more family of substitutions, namely, when
the chiral fields' dependence on $z$ and $t$ is determined by two functional
parameters $\xi$ and $\eta$, satisfying the equations
\beq\label{Eqd}
    \xi_z=\eta_t, \cm      \xi_t = \eta_z
\eeq
and
\bearr\label{Sol2}
   \gt=\gt(\xi,\eta),   \qquad  \chi=\chi(\xi,\eta), \nnn
   \phi=\phi(\xi,\eta),  \qquad    \psi=\const.
\ear
A solution to (\ref{NSMequat}) will be found in the form (\ref{Sol1}).
As follows from (\ref{Eqd}),
\beq\label{Solxid}
  \DAL\xi=0, \qquad \DAL\eta=0, \qquad  \xi_z\eta_z-\xi_t\eta_t=0.
\eeq
It leads to
\bear\label{Solxit}
  \xi(z,t) \eql g(z+t)+f(z-t), \nn
      \eta \eql g(z+t)-f(z-t).
\ear

It is possible to obtain a set of differential equations
for $\gt(\xi,\eta)$, $\chi(\xi,\eta)$ and $\phi(\xi,\eta)$
by substituting (\ref{Sol2}) to \eqs (\ref{NSMequat}) with (\ref{Solxid}):
\bear\label{PDUxi}
  \ddot\gt-\frac{1}{2}\dot\chi^2\sh2\gt  \eql 0,\nn
  \ddot\chi+{2}\dot\gt\dot\chi\cth\gt    \eql 0,\nn
  \ddot\phi-\frac{1}{2}\dot\gt^2-\frac{1}{4}\dot\chi^2\sh^2\gt \eql 0,\yy
\label{PDUeta}
  \gt''-\frac{1}{2}\chi'^2\sh2\gt  \eql 0,\nn
        \chi''+{2}\chi'\gt'\cth\gt \eql 0,\nn
  \phi''-\frac{1}{2}\gt'^2-\frac{1}{4}\chi'^2\sh^2\gt \eql 0.
\ear
Here
\[
     \phi'_a=\frac{d}{d\eta}\phi_a, \qquad
     \phi''_a=\frac{d^2}{d\eta^2}\phi_a,\ldots, \qquad a=1,2,3,4.
\]
\eqs (\ref{PDUxi}) and (\ref{PDUeta}) may be integrated independently over
$\xi$ and $\eta$ and the resulting solutions have the same form as
(\ref{SolF1}).  But in this case the real functions
$\phi_0$, $\phi_1$, $\xi_0$, $\chi_0$, $k$, $c$ depend on $\xi$ for
(\ref{PDUeta}) and depend on $\eta$ for (\ref{PDUxi}).

The symmetry of the solutions with respect to the substitution of $\xi$ to
$\eta$ makes it possible to write the simplest form of the solutions,
satisfying the continuity condition for the dependence of $\phi_a$ on
the parameters $\xi$ and $\eta$. To perform it in (\ref{SolF1}), it
is necessary to use
\bearr
  k=k_0(\eta-\eta_0), ~á=c_0(\eta-\eta_0), ~\phi_1=p_0(\eta-\eta_0),\nnn
  \phi_0,\ \chi_0,\ c_0,\ k_0,\ p_0,\ \eta_0,\ \xi_0=\const.
\earn
We should note that the solution obtained for a given nonlinear
$\gs$ model is a trivial generalization of the one-parametric
solution (\ref{SolF1}).
The arbitrary functions $f(z-t)$ and $g(z+t)$ in (\ref{Solxit}) allow one to
make the substitution
\bearr\label{}
  \xi = f(z-t)+g(z+t) \to \xi=F(z-t)+G(z+t),\nnn
   F(z-t)=f^2(z-t)-(\xi_0+\eta_0)f(z-t),    \nnn
   G(z+t)=-g^2(z+t)+(\xi_0-\eta_0)g(z+t),
\ear
which leads to a coincidence between (\ref{Solxit}) and (\ref{SolF1}).

The problem of a two-parameter nontrivial solution for the above $\gs$ model
is still open.

The problem may be solved by studying the equations for the
functions $\phi_0$, $\phi_1$, $\xi_0$, $\chi_0$, $k$ ,$c$ depending on $\xi$
and $\eta$, following from the coincidence of the solutions for the fields
$\phi_0$, $\phi_1$, $\xi_0$, $\chi_0$, $k$, $c$ satisfying (\ref{PDUxi})
and (\ref{PDUeta}). This problem is perhaps connected with the chiral space
structure and symmetries.  It may play an important role in the analysis of
the nonlinear $\gs$ model for other chiral space metrics.

\section{Conclusions}

 The main result of the present paper may be formulated as follows.

   The isometric ansatz method, derived from a symmetry investigation,
   allows one to obtain an exact solution for the chiral $\gs$ model of
   plane-symmetric gravitational vacuum spaces. Neither the inverse
   scattering problem method \cite{belzak78}, nor the functional parameter
   method described here allow one to obtain such a solution.  But the
   isometric ansatz cannot cover the whole family of possible solutions for
   the given $\gs$ model, as in the case of monopole family solutions [8].

   A new family of solutions for the effective nonlinear chiral $\gs$ model
   of plane-symmetric gravitation vacuum spaces is obtained.

   The explicit form of the solution yielded by the functional parameter
   method is obtained in a simpler way than using the
   Belinski-Zakharov method based on the ISPM. The solutions obtained may
   be used for formulating initial and boundary-value problems for the
   gravitational field of plane-symmetric vacuum spaces due to a very wide
   family of functions in the solution constructed by this method.

   The functional parameter method is equivalent to the B\"acklund
   transformation of the d'Alembert equation to the gravitational
   field equations for plane-symmetric space-times. This equivalence may be
   used for obtaining new exact solutions for nonlinear $\gs$
   models corresponding to other classes of chiral space metrics.

\Acknow{This work was carried out under partial financial support of the
Centre of CosmoParticle Physics ``Cosmion''.}

\end{document}